\documentclass[aps,prd,amsmath,nofootinbib,preprint,tightenlines,showpacs]{revtex4}
\newcommand{\be}{\begin{equation}}
\newcommand{\ee}{\end{equation}}
\newcommand{\bea}{\begin{eqnarray}}
\newcommand{\eea}{\end{eqnarray}}

\def\bA{\mathbf{A}}
\def\bB{\mathbf{B}}
\def\bE{\mathbf{E}}
\def\bK{\mathbf{K}}

\def\bV{\mathbf{V}}
\def\bX{\mathbf{X}}
\def\bY{\mathbf{Y}}
\def\bZ{\mathbf{Z}}
\def\order{O}

\def\calM{\mathcal{M}}

\def\Fermi{\mathop{\rm Fermi}\nolimits}

\usepackage{epsfig}

\begin{document}

\title{Multi-step Fermi normal coordinates}

\author{Eleni-Alexandra Kontou}
\author{Ken D. Olum}
\affiliation{Institute of Cosmology, Department of Physics and Astronomy,\\ 
Tufts University, Medford, MA 02155, USA}

\begin{abstract}

We generalize the concept of Fermi normal coordinates adapted to a
geodesic to the case where the tangent space to the manifold at the
base point is decomposed into a direct product of an arbitrary number
of subspaces, so that we follow several geodesics in turn to find the
point with given coordinates.  We compute the connection and the
metric as integrals of the Riemann tensor.  In the case of one
subspace (Riemann normal coordinates) or two subspaces, we recover
some results previously found by Nesterov, using somewhat different
techniques.

\end{abstract}

\pacs{02.40.Ky 	
      04.20.Cv  
}

\maketitle

\section{Introduction}

The construction of Riemann normal coordinates is well known.  For any
point $p$ of a Riemannian or Lorentzian manifold $(\calM, g)$ and any
vector $\bV_p$ at $p$ there exists a maximal geodesic
$\gamma_\bV(\lambda)$ with starting point $p$ and initial direction
$\bV_p$. We define the exponential map $\exp_p$ that takes a subset of
$T_p$, the tangent space to $\calM$ at $p$, into $\calM$, such that
$\exp_{p}(\bV$) is the point $q$ a unit parameter distance along the
geodesic $\gamma_\bV$ from $p$.

We can choose an orthonormal tetrad basis $\{ \bE_{(\alpha)} \}$ at
$p$ and then define the coordinates at $q$ by the relation
$q=\exp(x^\alpha\bE_{(\alpha)})$. Such coordinates are called
Riemann normal coordinates.

The Fermi normal coordinate construction \cite{Manasse:1963zz} is also
well known.  We start with a timelike geodesic $\gamma_\bK(\lambda)$
with tangent vector $\bK$ at $p$.  (We will consider only geodesics,
not arbitrary timelike curves.)  Given any vector $\bV\in T_p(\calM)$,
we can write $\bV=\bA+\bB$, where $\bA$ is in the direction of $\bK$
and $\bB$ perpendicular to $\bK$.  We then let $q = \exp_p(A)$ and
define a map $\Fermi_p$ such that $\Fermi_p(V) = \exp_q(B)$.  That is
to say, $\Fermi_p(V)$ is found by first moving unit distance along the
geodesic $\gamma_\bA$ from $p$ to $q$.  We parallel transport $B$ from
$p$ to $q$ and then move unit distance along the geodesic whose
tangent vector at $q$ is $B$.

We can define an orthonormal basis $\{\bE_{(\alpha)}\}$ at $p$
such that $\bE_{(0)}$ is parallel to $\bK$.  Then $\bA =
x^{0}\bE_{(0)}$, $\bB = x^i\bE_{(i)}$, giving the usual construction
of Fermi normal coordinates \cite{Manasse:1963zz}.

In this paper, we will generalize this construction to allow an
arbitrary number of arbitrary subspaces, rather than just a timelike
geodesic and the perpendicular space, and an arbitrary number $d$ of
dimensions.  In Sec.~\ref{sec:Fermi} we will construct the generalized
coordinate system, in Sec.~\ref{sec:connection} we compute the
connection, and in Sec.~\ref{sec:metric} we compute the metric in
the generalized Fermi coordinates.  We conclude in
Sec.~\ref{sec:conclusion}.

We use the sign convention $(+,+,+)$ in the classification of Misner,
Thorne and Wheeler \cite{MTW}.

\section{Multi-step Fermi coordinates}\label{sec:Fermi}

Consider a $d$-dimensional Riemannian or Lorentzian manifold $(\calM,
g)$.  We will start our construction by choosing a base point
$p\in\calM$.  We decompose the tangent space $T_p$ into $n$ subspaces,
$T_p = A^{(1)}_p \times A^{(2)}_p \times A^{(3)}_p \ldots \times
A_p^{(n)}$ so that any $\bV\in T_p$ can be uniquely written as $\bV=
\bV_{(1)} + \bV_{(2)} + \bV_{(3)} + \ldots + \bV_{(n)}$.  We choose,
as a basis for $T_p$, $d$ linearly independent vectors
$\{\bE_{(\alpha)}\}$ adapted to the decomposition of $T_p$ so that for
each $m=1\ldots n$, $\{\bE_{(\alpha)}|\alpha\in c_m\}$ is a basis for
$A_p^{(m)}$, where ${c_1,c_2, \ldots c_n}$ is an ordered partition of
$\{1\ldots d\}$. Thus each $V_{(m)}=\sum_{\alpha \in c_m}
x^\alpha\bE_{(\alpha)}$. The vectors $\{\bE_{(\alpha)}\}$ need not be
normalized or orthogonal.

\begin{figure}
\epsfysize=38mm
\epsfbox{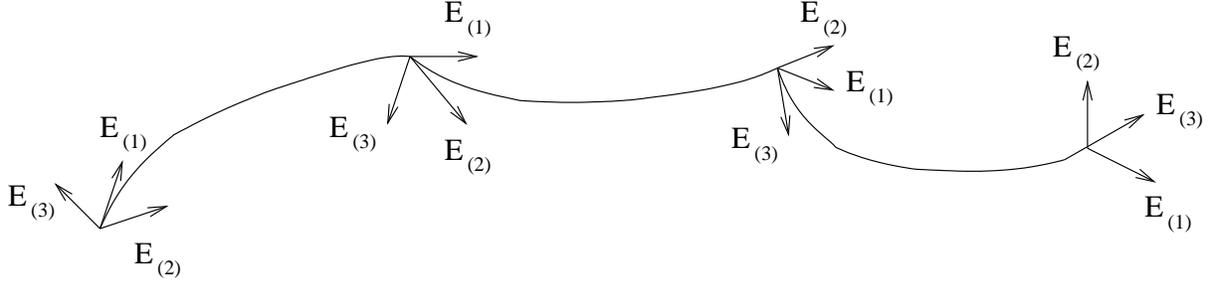}
\caption{Construction of 3-step Fermi coordinates in 3 dimensions.  We
  travel first in the direction of $\bE_{(1)}$, then $\bE_{(2)}$, then
  $\bE_{(3)}$, parallel transporting the triad as we go.}
\label{fig:3step}
\end{figure}

The point corresponding to coordinates $x^a$ is then found by starting
from $p$ and going along the geodesic whose whose tangent vector is
$\bV_{(1)}$, parallel transporting the rest of the vectors, then along
the geodesic whose tangent vector is $\bV_{(2)}$, and so on.  An
example is shown in Fig.~\ref{fig:3step}.
With that general construction of multi-step Fermi coordinates we can
define a general Fermi mapping $q=\Fermi_p(\bV)$ given by
\bea
q_{(0)}&=&p \nonumber\\
q_{(1)}&=&\exp_p(\bV_{(1)}) \nonumber\\
q_{(2)}&=&\exp_{q_{(1)}}(\bV_{(2)}) \\
\dots \nonumber\\
q=q_{(n)}&=&\exp_{q_{(n-1)}}(\bV_{(n)})\nonumber
\eea
From that general construction we can return to the original Fermi
case by choosing $c_1=\{t\}$ and $c_2=\{x,y,z\}$.  In the Lorentzian
case, we could also choose a pseudo-orthonormal tetrad
${\bE_u, \bE_v, \bE_x, \bE_y}$, with $\bE_u$ and $\bE_v$ null, $\bE_u
\cdot \bE_v = -1$, and other inner products vanishing, and $c_0=\{u\}$
and $c_1=\{v,x,y\}$.

For later use we will define
\bea
\bV_{(\leq m)}&=&\sum_{\alpha\in c_1 \cup c_2 \cup \dots \cup c_m} x^\alpha
\bE_{(\alpha)} = \sum_{l=1}^m \bV_{(l)}\\
\bV_{(<m)}&=&\bV_{(\le (m-1))} = \bV_{(\le m))} - \bV_{(m)}
\eea
Then we can write $q_{(m)} = \Fermi_p(\bV_{(\leq m)})$.

An example of a spacetime where these multi-step Fermi coordinates
might be used comes from brane-world models.  A general brane-world
metric with one extra dimension is
\be
ds^2=b(w) [-dt^2+a(t)(dx^2+dy^2+dz^2)]+dw^2.
\ee
A simpler metric of that form is used, for example, in
\cite{hep-ph/9905221}. In this kind of spacetime it might be useful to
introduce three-step Fermi coordinates with: $c_1=\{w\}$, $c_2=\{t\}$
and $c_3=\{x,y,z\}$.

\section{Connection}\label{sec:connection}

We will parallel transport our orthonormal basis vectors
$\bE_{(\alpha)}$ along the geodesics that generate the coordinates,
and use them as a basis for vectors and tensors throughout the region
of $\calM$ covered by our coordinates.  Components in this basis will
be denoted by Greek indices.  We will use Latin letters from the
beginning of the alphabet to denote indices in the Fermi coordinate
basis.  Of course at $p$, there is no difference between these bases.

Latin letters from the middle of the alphabet will denote the
subspaces of $T_p$ or equivalently the steps of the Fermi mapping
process.

We would like to calculate the covariant derivatives of the basis
vectors, $\nabla_\beta\bE_{(\alpha)}$.  Having done so, we can
calculate the covariant derivative of any vector field $\bV =
V^\beta \bE_{(\beta)}$ along a curve $f(\lambda)$ as
\be \label{eqn:covariant}
\frac{DV^\beta}{d\lambda} =\frac{dV^\beta}{d\lambda}  
+ V^\gamma \left(\frac{\partial}{\partial\lambda}\right)^\alpha
 \nabla_\alpha E_{(\gamma)}^\beta
\ee

To evaluate $\nabla_\beta\bE_{(\alpha)}$ at some point
$q_1=\exp_p(\bX)$, consider an infinitesimally separated point
$q_2=\exp_p(\bX+\bE_{(\beta)} dx)$.  The covariant derivative of
$\bE_{(\alpha)}$ at $q_1$ is the difference between
$\bE_{(\alpha)}(q_2)$ parallel transported to $q_1$ and the actual
$\bE_{(\alpha)}(q_1)$, divided by $dx$.  That difference is the same as
the change in $\bE_{(\alpha)}$ by parallel transport around a loop
following the geodesics from $q_1$ backward to $p$, the
infinitesimally different geodesics forward from $p$ to $q_2$, and the
infinitesimal distance back to $q_1$.  We can write this loop parallel
transport as an integral over the Riemann tensor.

\begin{figure}
\epsfysize=60mm
\epsfbox{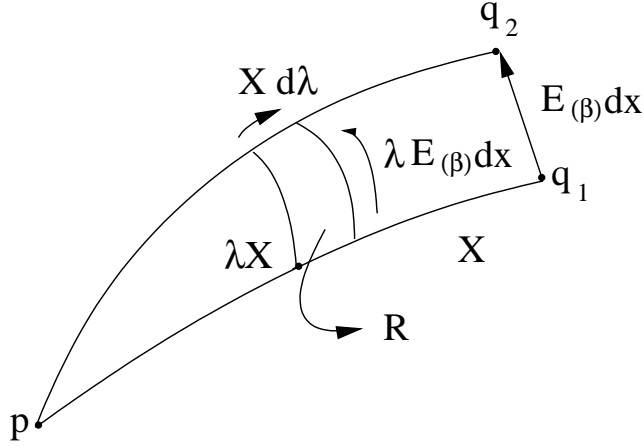}
\caption{The covariant derivative $\nabla_\beta \bE_{(\alpha)}$ is the
change in $\bE_{(\alpha)}$ under parallel transport along the path
$q_1\to p \to q_2 \to q_1$, divided by $dx$.  The parallel transport can
be decomposed into a series of transports clockwise around trapezoidal
regions with sides $\lambda \bE_{(\beta)} dx$ and $\bX d\lambda$.}
\label{fig:Riemann}
\end{figure}

Let us first consider the Riemannian case, as shown in
Fig.~\ref{fig:Riemann}.  The total parallel transport can be
written as the sum of parallel transport around a succession of small
trapezoidal regions whose sides are $\lambda \bE_{(\beta)}$ and $\bX
d\lambda$.   By using the definition of the Riemann tensor we have
\be\label{eqn:Riemanncovariant}
\nabla_\beta E_{(\alpha)}^\gamma = \int_0^1
d\lambda\, {R^\gamma}_{\alpha \delta \beta}(\lambda \bX) \lambda X^\delta\,.
\ee
Here $R$ is evaluated at the point $\exp_p(\lambda\bX)$, which we
have denoted merely $\lambda\bX$ for compactness.

Equation~(\ref{eqn:Riemanncovariant}) reproduces Eq.~(13) of
Ref.~\cite{Nesterov:1999ix}.  Note, however, that
Eq.~(\ref{eqn:Riemanncovariant}) is exact and does not require $R$ to
be smooth, whereas that of Ref.~\cite{Nesterov:1999ix} was given as
first order in $R$ and was derived by means of a Taylor series.

We see immediately that the covariant derivative of any $\bE_{(\alpha)}$
at $\bX$ in the direction of $\bX$ vanishes.  This
happens simply because changes with $d\bX$ in the direction of
$\bX$ correspond to additional parallel transport of
$\bE_{(\alpha)}$.

\begin{figure}
\epsfysize=40mm
\epsfbox{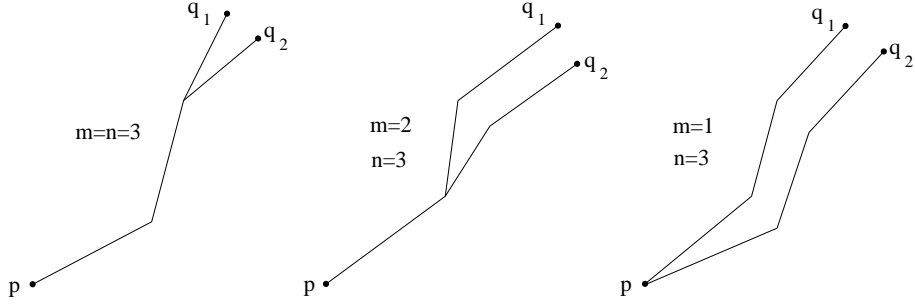}
\caption{Original and displaced geodesics for Fermi coordinates with
  $n=3$ and $m=3$, 2, and 1.}
\label{fig:fermipaths}
\end{figure}

Let us now consider the general case where there are $n$ steps, and
compute $\nabla_\beta \bE_{(\alpha)}$.  Since the coordinates are
adapted to our construction, the index $\beta$ must be in some
specific set $c_m$, which is to say that the direction of the
covariant derivative, $\bE_{(\beta)}$, is part of step $m$ in the
Fermi coordinate process.  We will write the function that gives that
$m$ as $m (\beta)$.  Some particular cases are shown in
Fig.~\ref{fig:fermipaths}.

If $m=n$ (leftmost in Fig.~\ref{fig:fermipaths}), only the last step
is modified.  The integration is exactly as shown in
Fig.~\ref{fig:Riemann}, except that it covers only the final geodesic
from $\bX_{(<n)}$ to $\bX_{(n)}$,
\be\label{eqn:covariantlast}
\nabla_\beta E_{(\alpha)}^\gamma= \int_0^1 d\lambda\, 
 {R^\gamma}_{ \alpha \delta \beta}(\bX_{(<n)} + \lambda \bX_{(n)})
 \lambda X_{(n)}^\delta.
\ee

\begin{figure}
\epsfysize=60mm
\epsfbox{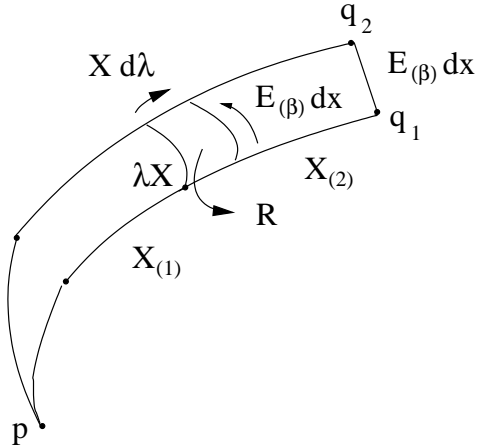}
\caption{Part of the calculation of $\nabla_\beta
  E_{(\alpha)}$ in the case $n=2$, $m(\beta)=1$.  The geodesic of the
  first step has been modified, causing the geodesic in the second
  step to be displaced.  The parallel transport integrates the Riemann
  tensor over a series of rectangular regions between the 2
  second-step geodesics.}
\label{fig:Fermi}
\end{figure}

If $m<n$, then we are modifying some intermediate step, and the path
followed at later steps is displaced parallel to itself.  In that case
we get an integral over rectangular rather than trapezoidal regions,
as shown in  Fig.~\ref{fig:Fermi}.
For general $m$ there is a contribution for each step $j \ge m$.  The
$j=m$ contribution integrates over trapezoids that grow with
$\lambda$, while the $j>m$ contributions integrate over rectangles
with fixed width $dx$.  We can write the complete result
\be \label{eqn:completecovariant}
\nabla_\beta E^\gamma_{(\alpha)}=\sum_{j=m}^n 
\int_0^1 d\lambda\, a_{jm}(\lambda)
{R^\gamma}_{ \alpha \delta \beta}(\bX_{(< j)} + \lambda \bX_{(j)}) X_{(j)}^\delta
\ee
where $m = m(\beta)$ and
\be
a_{jm}(\lambda)=\begin{cases}
1 & j \neq m\\
\lambda & j=m\,.
\end{cases}
\ee
Equation (\ref{eqn:completecovariant}) is exact and
includes Eqs.~(\ref{eqn:Riemanncovariant},\ref{eqn:covariantlast}) as
special cases.

Consider the case where $c_1$ consists only of one index.  If $m>1$,
there is no $j=1$ term in Eq.~(\ref{eqn:completecovariant}).  If
$m=1$, then $\beta$ is the single index in $c_1$, and
$X_{(1)}^\delta= 0$ unless $\delta =\beta$, so the $j=1$ term in
vanishes because $R_{\alpha\gamma\delta\beta}$ is antisymmetric
under $\delta \leftrightarrow\beta$.  Thus there is never a $j=1$
contribution to Eq.~(\ref{eqn:completecovariant}) when there is only
one index in $c_1$.

Now suppose $X$ lies on the first generating geodesic, so $X_{(j)} =
0$ for $j>1$.  Then all $j>1$ terms vanish in
Eq.~(\ref{eqn:completecovariant}).  So if $c_1$ consists only of one
index, all Christoffel symbols vanish at $X$.  This is well known in
the case of the usual Fermi coordinates.

\section{Metric}\label{sec:metric}

Now we would like to compute the metric $g$ at some point $\bX$.
Specifically, we would like to compute the metric component $g_{ab}$
in our generalized Fermi coordinates.

We will start by considering the vectors $\bZ_{(a)} =
\partial/\partial x^a$.  These are the basis vectors of the Fermi
coordinate basis for the tangent space, so the metric is given by
$g_{ab} =\bZ_{(a)} \cdot \bZ_{(b)}$.  Thus if we compute the
orthonormal basis components $Z_{(a)}^\alpha$ we can write
$g_{ab}=\eta_{\alpha \beta} Z_{(a)}^\alpha Z_{(b)}^\beta$.

Again we will start with the case of Riemann normal coordinates.  Let
$W(t,s)$ be the point $\exp_p s(\bX+t\bE_{(a)})$.  Define $\bY =
\partial W/\partial t$ and $\bV = \partial W/\partial s$.  Then 
$\bY(X) = \bZ_{(a)}$ and $V^\beta = X^\beta + t\delta^\beta_a$.
The components of $\bZ_{(a)}$ at $\bX$ can be calculated by integration,
\be\label{eqn:Yintegral}
Z_{(a)}^\beta(\bX)=Y^\beta(\bX)=\int_0^1 ds
\frac{\partial Y^\beta(s \bX)}{\partial s} .
\ee
Because the orthonormal basis is parallel transported we can write
\be\label{eqn:dYDV}
\frac{d}{ds} Y^\beta=\frac{D Y^\beta}{ds}.
\ee
By construction, the Lie
derivative $L_\bV\bY = 0$ and thus \cite[Ch.~4]{HawkingEllis}
\be\label{eqn:DYDV}
\frac{D\bY}{ds} = \frac{D\bV}{dt}
\ee
From Eq.~(\ref{eqn:covariant}) we have
\be\label{eqn:Vcovariant}
\frac{DV^\beta}{dt}=\frac{dV^\beta}{dt}
+V^\gamma Y^\alpha \nabla_\alpha E_{(\gamma)}^\beta 
= \delta_a^\beta+s \delta_a^\alpha V^\gamma \nabla_\alpha
E_{(\gamma)}^\beta + \order(R^2).
\ee
where we have retained $\delta_a^\alpha$ instead of writing $\nabla_a
E_{(\gamma)}^\beta$ to make it clear
that the covariant derivative is with respect to the orthonormal
basis.

From Eq.~(\ref{eqn:Riemanncovariant}) we have
\be\label{eqn:covariantE}
\nabla_\alpha E_{(\gamma)}^\beta( s \bX)=
\int_0^1 d\lambda\,\lambda {R^\beta}_{\gamma \delta \alpha}(\lambda s \bX)  sX^\delta
=\frac{1}{s}\int_0^s d \lambda\, \lambda {R^\beta}_{\gamma \delta \alpha}(\lambda  \bX)X^\delta
\ee
Taking $t=0$, $\bV$ is just $\bX$. Combining Eqs.~(\ref{eqn:Yintegral}-\ref{eqn:covariantE}), we
find
\bea\label{eqn:ZRiemann}
Z_{(a)}^\beta(\bX)&=& \int_0^1 ds \left[\delta_a^\beta
+ \delta_a^\alpha \int_0^s d \lambda\, \lambda {R^\beta}_{\gamma \delta
  \alpha}(\lambda  \bX)X^\delta X^\gamma\right]+\order(R^2)\nonumber\\
 &=&\delta_a^\beta+\delta_a^\alpha\int_0^1  d\lambda\,
\lambda (1-\lambda) {R^\beta}_{\gamma \delta \alpha}(\lambda \bX)X^\delta X^\gamma+\order(R^2).
\eea

From Eq.~(\ref{eqn:ZRiemann}), the metric is given by
\be\label{eqn:Riemannmetric}
g_{ab}=\eta_{ab}+2 \delta_a^\alpha \delta_b^\beta \int_0^1  d\lambda\, \lambda (1-\lambda) R_{\alpha \gamma \delta \beta}(\lambda \bX)X^\delta X^\gamma+\order(R^2).
\ee
Equation~(\ref{eqn:Riemannmetric}) reproduces Eq.~(14) of
Ref.~\cite{Nesterov:1999ix}\footnote{Ref.~\cite{Nesterov:1999ix} uses
  the same sign convention for ${R^\alpha}_{\beta\gamma\delta}$ as the
  present paper, but the opposite convention for $g_{ab}$ and
  consequently also for $R_{\alpha\beta\gamma\delta}$.}.

Next let us consider the case where there are $n$ steps in our
procedure.  We will define a set of functions $W_j$ as
\be
W_j(s)=\Fermi_p(\bX_{(<j)}+s\bX_{(j)}).
\ee
The path $W_j(s), j=1\ldots n, s=0\ldots 1$ traces the geodesics
generating the Fermi coordinates for the point $X$.
Now consider $\bZ_{(a)} = \partial/\partial x^a$.  Let $m=m(a)$,
so $\bZ_{(a)}(p)\in A_p^{(m)}$.  Then let
\be
W_j(s,t)=\Fermi_p \begin{cases}
s \bX_{(j)} & j<m\\
\bX_{(<j)}+s(\bX_{(j)}+t\bE_{(a)}) & j=m\\
\bX_{(<j)}+t\bE_{(a)}+s\bX_{(j)} & j>m 
\end{cases}
\ee
Let $\bY = \partial W/\partial t$ and $\bV = \partial W/\partial
s$ as before.
To find $\bZ_{(a)}$ we now must integrate over a multi-step path from $p$,
\be\label{eqn:YintegralFermi}
Z_{(a)}^\beta(\bX)=\sum_{j=1}^n \int_0^1 ds \frac{\partial
Y^\beta(W_j(s))}{\partial s}.
\ee
The generalized version of Eq.~(\ref{eqn:Vcovariant}) is
\be\label{eqn:VcovariantFermi}
\frac{DV^\beta (W_j(s,t))}{dt}=\frac{dV^\beta}{dt}
+V^\gamma Y^\alpha \nabla_\alpha E_{(\gamma)}^\beta 
=\begin{cases}
0 & j<m\\
\delta_a^\beta+s \delta_a^\alpha V^\gamma \nabla_\alpha
E_{(\gamma)}^\beta + \order(R^2) & j=m\\
\delta_a^\alpha V^\gamma \nabla_\alpha
E_{(\gamma)}^\beta + \order(R^2) & j>m.
\end{cases}
\ee 
Now
\be\label{eqn:covariantEfermi}
\nabla_\alpha E_{(\gamma)}^\beta(W_j(s))=
\sum_{k=m}^{j} \frac{1}{s_{kj}(s)}  \int_0^{s_{kj}(s)} d\lambda\, a_{km}(\lambda) {R^\beta}_{\gamma \delta \alpha}(\bX_{(<k)}
+\lambda\bX_{(k)}) X_{(k)}^\delta
\ee
where
\be
s_{kj}(s)=\begin{cases}
1 & k \neq j \\
s & k=j.
\end{cases}
\ee
The $k = j$ term is analogous to Eq.~(\ref{eqn:covariantE}), while
the others have no dependence on $s$.

Combining
Eqs.~(\ref{eqn:dYDV},\ref{eqn:DYDV},\ref{eqn:YintegralFermi}--\ref{eqn:covariantEfermi})
we get
\be
Z_{(a)}^\beta(\bX)= \delta_a^\beta+ F_a^\beta+ \order(R^2)
\ee
where
\bea
F_\alpha^\beta&=&\sum_{j=m}^n \sum_{k=m}^{j} 
\int_0^1 ds \int_0^{s_{kj}(s)} d\lambda\, a_{km}(\lambda) 
{R^\beta}_{\gamma \delta \alpha}(\bX_{(<k)} + \lambda \bX_{(k)}) X_{(k)}^\delta X_{(j)}^\gamma\nonumber\\
&&= \sum_{j=m}^n \sum_{k=m}^{j} \int_0^1 d\lambda\,a_{km}(\lambda)b_{kj}(\lambda) 
{R^\beta}_{\gamma \delta \alpha}(\bX_{(<k)} + \lambda \bX_{(k)}) X_{(k)}^\delta X_{(j)}^\gamma
\eea
where $m = m(\alpha)$ and
\be
b_{kj}(\lambda)= \begin{cases}
1 & k \neq j\\
1-\lambda & k=j
\end{cases}
\ee
So the metric is
\be\label{eqn:smetric}
g_{ab}=\eta_{\alpha \beta} Z_{(a)}^\alpha Z_{(b)}^\beta
= \eta_{ab}+F_{ab}+F_{ba}+\mathcal{O}(R^2)
\ee
where
\be\label{eqn:Flower}
F_{\alpha\beta}= \sum_{j=m}^n \sum_{k=m}^{j} \int_0^1 d\lambda\,
a_{km}(\lambda)b_{kj}(\lambda) R_{ \alpha\gamma\delta\beta}(\bX_{(>k)} + \lambda \bX_{(k)}) X_{(k)}^\delta
X_{(j)}^\gamma
\ee
where $m = m(\beta)$.

Once again consider the case where $c_1$ contains only one index.  As
discussed with respect to Eq.~(\ref{eqn:completecovariant}), if
$\beta\in c_1$, there is no nonvanishing $k=1$ term in
Eq.~(\ref{eqn:Flower}).  Thus $g_{ab} = \eta_{ab}$ at points on the
first generating geodesic.  This is also well known in the usual Fermi
case.

Now suppose $c_1$ consists only of one index and furthermore $n = 2$.
The only possible term in Eq.~(\ref{eqn:Flower}) is then
$j=k=2$, so
\be\label{eqn:Flower2}
F_{\alpha\beta}= \int_0^1 d\lambda\,
a_{2m}(\lambda)(1-\lambda) R_{ \alpha\gamma\delta\beta}(\bX_{(1)} + \lambda \bX_{(2)}) X_{(2)}^\delta
X_{(2)}^\gamma.
\ee
where $m = m(\beta)$.
Equation Eq.~(\ref{eqn:Flower2}) is equivalent to Eq.~(28) in
Ref.~\cite{Nesterov:1999ix} in the case where the generating curve of
the Fermi coordinates is a geodesic.

Now we are in a position to discuss the region of the manifold
over which the multi-step Fermi coordinates are well defined.
Assuming there are no singularities or edges in the manifold, the only
problem would be if the metric $g_{ab}$ is degenerate, which in turn
can happen only if $F_{ab}$ is of order 1.  Thus the Fermi coordinates
will be well defined providing that \cite{Nesterov:1999ix}
\be\label{eqn:Rlimit}
|R_{ \alpha\gamma\beta\delta}| (X^\epsilon)^2 \ll 1
\ee
throughout the region of interest, for all
$\alpha,\gamma,\beta,\delta,\epsilon$.

In the case where there is only one index in $c_1$, there is no
contribution to $F_{ab}$ from $X_{(1)}$.  Then it is sufficient for
Eq.~(\ref{eqn:Rlimit}) to hold for $\epsilon>1$.  In other words, if the
first step is one-dimensional, it can be arbitrarily long
\cite{Manasse:1963zz}.

\section{Conclusion}\label{sec:conclusion}

We have generalized the usual Fermi normal coordinates in the case
where the generating curve is a geodesic to allow for any number of
steps and for a subspace of any dimension at each step.  We have
derived the connection (exactly) and the metric (to first order in the
curvature) as integrals over the Riemann tensor.  Our results
reproduce several formulas previously derived by Nesterov
\cite{Nesterov:1999ix} with a more geometric approach in a more
general setting, and without reference to any derivative of $R$.

\section*{Acknowledgments}

We would like to thank Jose Blanco and Ben Shlaer for helpful
discussions.  This research was supported in part by grant RFP3-1014
from The Foundational Questions Institute (fqxi.org).

\bibliography{no-slac,paper}

\end{document}